\documentclass[12pt]{article}

\newcommand{\be}{\begin{equation}}

\newcommand{\BM}{Bohmian mechanics}

\renewcommand{\c}{classical}

\newcommand{\cm}{classical motion}
\newcommand{\CM}{classical mechanics}
\newcommand{\cl}{classical limit}

\renewcommand{\d}{\delta}

\renewcommand{\div}{\mbox{\textup{\textrm{div}}}\,}

\newcommand{\ee}{\end{equation}}

\newcommand{\env}{environment}

\newcommand{\ex}{\textrm{e}}

\renewcommand{\Im}{\mbox{Im}}

\newcommand{\h}{\hbar}

\newcommand{\HJ}{Hamilton-Jacobi}

\newcommand{\NM}{Newtonian mechanics}

\newcommand{\ode}[2]{\frac{ d #1}{ d #2}}
\newcommand{\pd}{probability distribution}
\newcommand{\pde}[2]{\frac{\partial #1}{\partial #2}}

\newcommand{\q}{quantum}
\newcommand{\qf}{quantum force}

\newcommand{\QM}{quantum mechanics}
\newcommand{\qp}{quantum potential}

\renewcommand{\r}{\rho}

\newcommand{\Sc}{Schr\"{o}dinger}
\newcommand{\se}{Schr\"odinger's equation}

\newcommand{\trs}{trajectories}

\newcommand{\vel}{velocity}

\newcommand{\wf}{wave function}

\begin{document}

\title{
    What is Bohmian Mechanics}
\author{
    Valia Allori\thanks{email: allori@ge.infn.it}
     \and
     Nino Zangh\`{\i}\thanks{email: zanghi@ge.infn.it} 
     }
\date{\small Dipartimento di Fisica dell'Universit\`a di Genova\\
Istituto Nazionale di Fisica Nucleare, Sezione di Genova\\
via Dodecaneso 33, 16146 Genova, Italy}

\maketitle

\begin{abstract}
    Bohmian mechanics is a quantum theory with a clear ontology.
To make clear what we mean by this, we shall proceed by recalling
first what are the problems of \QM{}.  We shall then briefly sketch
the basics of \BM{} and indicate how Bohmian mechanics solves these
problems and clarifies the status and the role of of the quantum
formalism.
\end{abstract}

\section{What is quantum mechanics about?}
\label{subsec:The Problem of Quantum Mechanics}

The basic problem of quantum mechanics is that it is not clear what
quantum mechanics is about---what quantum mechanics describes---as
repeatedly stressed by John Bell \cite{bell}, and, more recently, by
Shelly Goldstein \cite{goldstein}.

It might seem that quantum mechanics is basically about the behavior
of wave functions.  However, as Scr\"{o}dinger has effectively shown
with his famous cat paradox \cite{schroedinger}, it turns out that the
complete state description cannot be given by the \wf{}---obeying
\se{}---as this would lead to paradoxical conclusions like, for
example, a superposition between a dead and an alive cat.  Bell has
rephrased this mental experiment in a less cruel way as follows:
consider a cat in a perfectly isolated room.  Together with the cat,
the experimenter has put in the room a radioactive source and a
complicated mechanics.  If a radioactive nucleus decays, the mechanism
opens a source of milk such that it fills a cup and the cat can drink.
The room has no window so that what happens inside is completely
hidden to the experimenter: she doesn't know whether the cat is still
hungry or if she enjoyed her meal.  In this way the radioactive decay,
a microscopic event, influences directly a macroscopic event, like the
presence or not of some milk molecules in the stomach of the cat.

{}From the mathematical rules of \QM{} it follows that, given that the
\wf{} of the radioactive nucleus is in a superposition of decayed--non
decayed \wf{}, the cat is nor hungry nor filled up but it is a
superposition of both states.  However, {}from ordinary experience, we
know that macroscopic object cannot be in such a superposition of
states with macroscopically disjoint supports, so somewhere, somehow,
\QM{} gives the wrong answer.  Note that, if the experimenter opens
the door of the room, finds out the cat always or in one or in the
other state: as a consequence of observation (measurement), the \wf{}
has collapsed in one of the two possibilities.

The Schr\"{o}dinger's cat paradox poses several questions: What is the
role of the observer?  Which observer is entitled to reduce the \wf{}?
Where is the border between the microscopic world, in which
superpositions can exist, and the macroscopic world, in which they
cannot?  Nobody has ever found satisfactory answers to these questions
within the standard framework of quantum mechanics.  Indeed, Bell has
drawn the conclusion that there are only two ways out: or to add
something to the \wf{} for the description of the state of the system,
or to modify \se{}.  While this latter path is that taken by the so
called theories of spontaneous localization, or shortly GRW theories
\cite{grw}, \BM{} is a theory that follows the first
direction.\footnote{\BM{} (also called ``pilot-wave theory''or
``causal interpretation'') was discovered in 1927 by De Broglie
\cite{solvay} and soon abandoned.  It was rediscovered, extensively
extended, and for the first time fully understood, in 1952 by David
Bohm \cite{bohm52}.  During the sixties, seventies and eighties, John
Bell was his principal proponent; his book \cite{bell} contains yet
unsurpassable introductions to \BM. Other standard references are the
books of Bohm and Hiley \cite{bohmhiley} and that of Holland
\cite{holland}.  The approach we are following here is that of the
``Rutgers-M\"unchen-Genova'' group (quite in line with the approach of
Bell), see, e.g., \cite{qe}, \cite{cushing}, \cite{goldstein},
\cite{shelly}, \cite{durr}.}

\section{Bohmian Mechanics}
\label{subsec:The State of a System and The Dynamical Laws}

The first step in the construction of a physical theory is to
establish what are the mathematical entities (particles, fields,
strings, \ldots) with which one intends to describe physical reality.
These mathematical entities are what the theory is about and they are
often called the ontology of the theory---a rather complicated way of
expressing a simple, even though deep, physical notion.

In nonrelativistic \BM{} the world is described by point--like
particles which follow trajectories determined by a law of motion.
The evolution of the positions of these particles is guided by the
\wf{} which itself evolves according to \se{}.  In other words, in
\BM{} the complete description of the state of an $N$ point--like
particle system is the couple $(\Psi,Q)$, where $\Psi=\Psi(q)$ and
$Q=( Q_1(t),...,Q_N(t))$ are respectively the \wf{} and the
\emph{actual} configuration of the system, with $Q_k$ denoting the
position of the $k$-th particle in ordinary three dimensional space.

One might think of \BM{} as a dynamical system and {}from this point
of view it can be compared with \CM{}.  While in \NM{} the dynamics of
the point particles is determined by a {\it second} order differential
equation
\be \frac{d^{2}Q_{t}}{d\,t^{2}}=\frac{1}{m}F(Q_t),
\label{eq:secondorder}
\ee in which $F(Q)$ is a force field, e.g,  derived {}from a potential
field $V$ as
$F(Q)=-\nabla V$, in \BM{} the point particles dynamics is given by a {\it
first} order differential equation
\be \ode{Q_t}{t} = v^{\Psi}(Q_t),
   \label{eq:bohmeq}
\ee where $v^{\Psi}=(v_1^{\Psi},...,v_N^{\Psi})$ is a \vel{} field on
the configuration space.  This fields is generated by the wave
function $\Psi$ which itself evolves according to \se{} \be
i\hbar\pde{\Psi}{t} = H\Psi \,,
    \label{eq:scevol}
\ee where $H$ is the Hamiltonian, e.g., given, for non relativistic
spinless particles, by
\begin{equation}
      H=-\sum_{k=1}^N\frac{\h^2}{2m_k}\nabla_k^2+V
      \label{eq:hamiltonian}
\end{equation}
The \vel{} field is determined by reasons of simplicity and symmetry \cite{qe},
\be
v_k^{\Psi}=\frac{\h}{m_k}
\Im\left [\frac{\nabla_k \Psi}{\Psi}\right ]
\label{eq:velofield}
\ee (with the wave function playing somehow the role of ``potential
field'' for the velocity field).  The factor $\frac{\h}{m}$ comes
{}from the requirement of Galilei invariance, the imaginary part is a
consequence of invariance for time reversal, the gradient is {}from
rotational invariance and the fact that one has to divide for $\Psi$
derives {}from the homogeneity of degree zero of the \vel{}
field---the fact that a quantum state is a ray in Hilbert space.  If
there is a magnetic field $B= {\rm rot} A$, $\nabla_k$ should
represent the covariant derivative associated with the vector
potential $A$.  If the \wf{} is a spinor, we should rewrite the \vel{}
field as \be v_k^{\Psi}=\frac{\h}{m_k} \Im\left [\frac{\Psi^*\nabla_k
\Psi}{\Psi^*\Psi}\right ],
\label{eq:velofieldspinor}
\ee where now in the numerator and denominator appears the scalar
product in the spinor space.

The global existence of Bohmian dynamics has been proven with full
mathematical rigor in \cite{brendl} where it has been shown that for a
large class of \Sc{} Hamiltonians (\ref{eq:hamiltonian}), including
Coulomb potential $V$ with arbitrary charges and masses, and
sufficiently regular initial datum $\psi_0$ of (\ref{eq:scevol}) the
solution of (\ref{eq:bohmeq}) exists uniquely and globally in time for
$| \psi_0|^2$-almost all initial configurations~$Q_0$.  \medskip

Equations (\ref{eq:bohmeq}) and (\ref{eq:scevol}) (together with
(\ref{eq:hamiltonian}) and (\ref{eq:velofield})) form a complete
specification of the theory.  Without any other axiom, all the
phenomena governed by nonrelativistic quantum mechanics, from spectral
lines and quantum interference experiments to scattering theory,
superconductivity and quantum computation follow {}from the analysis
of the dynamical system defined by (\ref{eq:bohmeq}) and
(\ref{eq:scevol}).

\section{Experimental Predictions}
\label{subsec:The Quantum Equilibrium Hypothesis and the Experimental
Predictions}

Bohmian mechanics makes the same predictions as does non relativistic
ordinary \QM{} for the results of any experiment, provided that we
assume a random distribution for the configuration of the system and
the apparatus at the beginning of the experiment given by
$\r(q,t)=|\Psi(q,t)|^2$.  In fact, consider the \q{} continuity
equation \be \pde{\r}{t} + \div J^{\Psi} =0,
\label{eq:conteq}
\ee which is, by himself, a simple consequence of \se{}.  Here $
J^{\Psi}=(J^{\Psi}_{1}, \ldots, J^{\Psi}_{N})$ is the \q{} probability
current \be J^{\Psi}_k= \frac{\hbar}{m_k} \Im \left [
\Psi^*\nabla_{\!k}\Psi\right ]= |\Psi|^2 v_k^{\Psi} .
\label{eq:current}
\ee Equation (\ref{eq:conteq}) becomes the \c{} continuity equation
\be \pde{\r}{t} + \div \r\, v^{\Psi} =0
     \label{eq:classconteq}
\ee for the system $dQ_t/dt=v^{\Psi}$ and it governs the evolution of
the probability density $\r$ under the motion defined by the guiding
equation (\ref{eq:bohmeq}) for the particular choice $\r=|\Psi|^2$.
In other words, if the probability density for the configuration
satisfies $\r(q,t_0)=|\Psi(q,t_0)|^2$ at some time $t_0$, then the
density to which this is carried by the motion (\ref{eq:conteq}) at
any time $t$ is also given by $\r(q,t)=|\Psi(q,t)|^2$.  This is an
extremely important property of any Bohmian system.  In fact it
expresses a compatibility between the two equations of motion defining
the dynamics, which we call {\it equivariance} of $|\Psi|^2$.

The above assumption, which guarantees agreement between \BM{} and
\QM{} regarding the results of any experiment, is what has been called
\cite{qe} the {\it {\q{} equilibrium hypothesis}}: when a system has a
\wf{} $\psi$, its configuration $Q$ is random with \pd{} given by \be
\r(q) =|\psi(q)|^2.
\label{eq:equivmeas}
\ee While the meaning and justification of this hypothesis---which
should be regarded as a local manifestation of a global {\it
quantum\/} equilibrium state of our universe---is a delicate matter,
which has been discussed at large elsewhere \cite{qe}, it is important
to recognize that, merely as a consequence of (\ref{eq:classconteq})
and (\ref{eq:equivmeas}), \BM{} is a counterexample to all of the
claims to the effect that a deterministic theory cannot account for
quantum randomness in the familiar statistical mechanical way, as
arising from averaging over ignorance: \BM{} is clearly a
deterministic theory, and, as we have just explained, it does account
for quantum randomness as arising from averaging over ignorance given
by $|\psi(q)|^2$.
\medskip

It is important to realize that non simply \BM{} makes the same
predictions as does orthodox quantum theory for the results of any
experiment, but that \emph{the quantum formalism of operators as
observables emerges naturally and simply from it as the very
expression of the empirical import of \BM{}}.  More precisely, it
turns out that in \BM{} self-adjoint operators arise in association
with \emph{specific} experiments as a tool to compactly express and
represent the relevant data---the results and their statistical
distributions---of these experiments.  The key ingredient to
understand how this comes about is to recall \cite{cushing} that a
completely general experiment is described by:
\begin{itemize}
\item a unitary map $U$ transforming the initial state of system and
apparatus $\psi_0(x)\otimes \phi_0(y)$ into a final state
$\Psi(x,y)=U\left( \psi_0(x)\otimes\phi_0(y)\right)$ ($x$ refers to
the configurations of the system and $y$ those of the apparatus);
\item a pointer variable $Z=F(Y)$ representing the pointer orientation
in terms of the microscopic configurations $Y$ of the apparatus.
\end{itemize}
It is a direct consequence of quantum equilibrium and linearity of
\se{} \cite{cushing}, that the \pd{} of the pointer variable $Z$ is a
measured--valued quadratic form on the Hilbert space of \wf{}, and, a
such, mathematically equivalent to a positive--operator--valued
measure (POVM).  It turn out that self--adjoint operators (which are,
by the spectral theorem, in one to one correspondence with
projector--valued measures) represent quantum observables associated
with the special class of repeatable experiments \cite{operator}.

Note that as a byproduct of the foregoing considerations one obtains a
very general notion of measurability: \be \mbox{
      \begin{minipage}{0.70\textwidth}
{\it A physical quantity is measurable only if its probability
distribution is a measure-valued quadratic form on the Hilbert space
of wave functions.}
  \end{minipage}
       }
\label{meas}
\ee
\bigskip

Sometimes it is claimed that it is possible to experimentally
discriminate between \BM{} and \QM{}.  This claim is however totally
unfounded: there {\it must} be experimental agreement as a consequence
of quantum equilibrium.  The experimental equivalence of \BM{} with
\QM{} might appear, somehow, a little frustrating fact: while, on one
hand, all the experimental evidence confirms \BM{} as well as \QM{},
on the other hand it would be easier it the experimental prediction
were different.  In fact, if there were a crucial experiment able to
discriminate between the two theories, there would be something {\it
objective} to establish which is the correct theory.  It must be made
clear, however, that the experimental equivalence of \BM{} with \QM{}
holds as long as the predictions of \QM{} are not ambiguous.  There
are in fact a variety of experimental issues that don't fit
comfortably within the standard operator quantum formalism, such as
dwell and tunneling times \cite{leavens}, escape times and escape
positions \cite{daumer}, scattering theory \cite{duemila}, but are
easily handled by Bohmian mechanics.

Actually, after the discussion of the previous sections, it should be
clear that the comparison shouldn't be made only on the level of
experimental prediction but, on the contrary, the decision of what is
the right theory should be taken on the deeper level of the ontology
of the theory---what the theory is about.

\section{The  Collapse of the wave function}
\label{subsec:The Wave Function of a Subsystem and the Collapse}

The existence of configurations in \BM{} allows for a natural and
clear notion of \wf{} of a subsystem.  In fact, consider a composite
system composed by a sub-system and by its \env{}.  If
$Q_t=(X_t,Y_t)$, where $X_t$ is the actual (i.e., what really is)
configuration of the sub-system at time $t$, and $Y_t$ is that its
\env{} at the same time, we can define the {\it {conditional}} \wf{}
for the $x$-system at time $t$ as \be \psi_t(x)=\Psi_t(x,Y_t),
\label{eq:conditional}
\ee that is, the \wf{} of the whole universe (the biggest system of
all) $\Psi_t$ calculated in the actual configuration of the \env{}.
Under appropriate conditions $\psi(x)$, satisfies \se{} in $x$.  In
this case it is indeed the {\it{effective}} \wf{} for the $x$-system,
that is, the collapsed \wf{} that the ordinary \q{} formalism assigns
to the subsystem after a \q{} measurement.  In fact, suppose $\Psi$
has the structure occurring in a measurement situation \be
\Psi_t(x,y)=\psi_t(x)\phi_t(y)+\Psi_t^{\perp}(x,y),
\label{eq:measurement}
\ee where $\phi_t(y)$ and $\Psi_t^{\perp}(x,y)$ (the part of $\Psi_t$
which is not $\psi_t(x)\phi_t(y)$) have macroscopically disjoint
$y$-supports.  If $Y_t$ belongs to the support of $\phi_t(y)$,
$\psi_t(x)$ is the effective \wf{} of the $x$-system at time $t$.
(For a clear exposition of this, see \cite{cushing} or \cite{qe}.)

Thus, the collapse of the \wf{} can be deduced {}from \BM{} without
introducing any active role to the observer.  Consider, again, the cat
paradox in the original version, were the two superposing states are
dead and alive cat.  In \BM{} at any time $t$ the cat is something
real, she is or dead or alive, independently on who is looking at her.
Note that she {\it can} be in a superposition state because the \wf{}
evolves according to \se{}, but in \BM{} the state of the system is
given by the couple $\left(\Psi,Q\right)$ of the \wf{} and the
configurations $Q=(Q_1,...,Q_n)$ of all the particles composing the
system (the cat).  Thus, according to which support $Q$ belongs to (to
those of the \wf{} $\Psi_{dead}$ describing the dead cat or to those
of the \wf{} $\Psi_{alive}$ describing the alive cat), the cat is
actually dead or alive.  Note that superpositions exist on all scales
({}from micro to macro) but don't influence at all the fact that the
cat is this or that.  At this point a question could arise: due to the
presence of a superposition \wf{}, could it be possible that the cat,
who at some time is dead, returns alive?  The cat has an actual
configuration, belonging (in our example) to the support of
$\Psi_{dead}$, and its evolution is guided by the \wf{}.  There seems
to be nothing to prevent to $Q$ to be guided in the support of
$\Psi_{alive}$, making the dead--alive transition possible.  Actually,
this is very unlikely to happen, in fact the supports of the two wave
functions are macroscopically distinguishable.  By this we mean that
the macroscopic variables, like, e.g., the temperature, assume
different values in the two states, even if the microscopic quantities
{}from which they have been derived might be similar.  The temperature
of an dead cat and of an alive cat are, in general, different.  Thus,
if $Q$ at some time belongs to the support of $\Psi_{dead}$, the
effect of $\Psi_{alive}$ is completely negligible: we can forget of it
for the dynamics of $Q$.  The dead--alive transition could be possible
if we would be able to bring the two wave functions close to each
other again.  But the probability of having success in this would be
even less probable than the fact that all the molecules of perfume we
have sprayed in a room would come back spontaneously in the
neighborhood of the bottle!

\section{Bohmian Mechanics and Newtonian Mechanics}
\label{subsec:Bohmian Mechanics and Newtonian Mechanics}
\label{subsec:Bohmian Mechanics and the Quantum Potential}

To point out some interesting features of \BM{}, it can be useful to
write the \wf{} $\Psi$ in the polar form \be \Psi=R\ex^{\frac{i}{\h}S}
\label{eq:polar}
\ee and then rewrite \se{} in terms of these new variables.  This is
indeed what Bohm originally did in his 1952 paper \cite{bohm52}.  In
this way one obtains {}from (\ref{eq:scevol}) a pair of coupled
equations: the continuity equation for $R^2$, \be \frac{\partial
R^2}{\partial t}+{\rm{div}}\left( \frac{\nabla_k S}{m}\right)R^2=0,
\label{eq:qcont}
\ee
   which suggests that $\r=R^2$ can
   be interpreted as
a probability density, and a modified \HJ{} equation for $S$
\be
\frac{\partial S}{\partial t}+\frac{(\nabla_k S)^2}{2m}+
V-\sum_k\frac{\h^2}{2m_k}\frac{\nabla_k^2 R}{R}=0
\label{eq:qHJ}
\ee
Note that this equation differs {}from the usual
classical \HJ{} equation
\be
\frac{\partial S}{\partial t}+\frac{(\nabla_k S)^2}{2m}+V=0
\label{eq:cHJ0}
\ee
   only by the appearance of an extra term,
   the {\it {\qp{}}}
   \be
   U\equiv-\sum_k\frac{\h^2}{2m_k}\frac{\nabla_k^2 R}{R}.
   \label{eq:potential}
\ee This modified \HJ{} equation can be used, together with the
continuity equation for $R$, to define particle \trs{} identifying the
velocity with $v_k=\frac{\nabla_k S}{m}$.  In this way the resulting
motion is precisely what would have been obtained classically if the
particles were subjected by the force generated by the \qp{} in
addition to the usual forces.

It should be noted, however, that the rewriting of \se{} through the
polar variables $(R,S)$ is somehow misleading.  In fact, first of all,
there is an increase in complexity: \se{} is a linear equation while
the modified \HJ{} equation is highly nonlinear and still requires the
continuity equation for its closure.  Note that, since in \BM{} the
dynamics is completely defined by \se{} (\ref{eq:scevol}) and the
guiding equation (\ref{eq:bohmeq}), there is no need of any further
{\it {axioms}} involving the \qp{} and thus it should not be regarded
as the most basic structure defining \BM{}.

\BM{} is not a rephrasing of \QM{} in classical terms.  It is not
simply \CM{} with an additional force term.  In \BM{} the velocities
are not independent of positions, as they are classically, but are
constrained by the guiding equation (\ref{eq:bohmeq}).  The correct
way of regarding to Bohmian mechanics is as a first-order theory, in
which the fundamental quantity is the position of particles, whose
dynamics is specified directly and simply by the velocity field
(\ref{eq:velofieldspinor}).  In \BM{} the second-order (Newtonian)
concepts of acceleration and force, work and energy play no
\emph{fundamental} role.  Rather, they are fundamental to the theory
to which \BM{} converges in the \cl{}, namely \NM{}.  In fact,
regardless of whether or not we think to the \qp{} as fundamental, it
can be useful: equation (\ref{eq:qHJ}), or, equivalently, \be
\frac{d^{2}Q_{t}}{d\,t^{2}} =-\nabla [V(Q_t)+
U(Q_t)],\label{eq:newtonequation} \ee show that all the {\it
{deviations}} {}from classicality are embodied in the \qf{} $-\nabla U
$, so that, whenever this force is negligible, there is \cm{}
\cite{holland}, \cite{bohmhiley}.  This observation is the starting
point for a complete derivation of the classical limit of quantum
mechanics. And the crucial step for such a derivation is to
characterize the physical conditions that guarantee smallness of the
\qf{} (see \cite{allori}, \cite{allori1}, \cite{allori2}).

\section{Nonlocality and Hidden Variables}
\label{subsec:NolLocality and Hidden Variables}

In the literature it is common to refer to \BM{} as a theory of {\it
hidden variables}.  This is a consequence of the famous EPR paper
\cite{einstein} in which Einstein, Podolsky and Rosen argued that
\QM{} might be incomplete.  Their proposal was to look for some non
measurable variables (somehow hidden) to complete the theory.

It should be stressed that the problems faced by Einstein, Podolsky
and Rosen in their paper was about the locality of quantum theory:
they assumed that reality is local, i.e. action at distance is
impossible, and proposed a mental experiment (that we shall not recall
here).  Their conclusions were that, if reality is local, \QM{} is
incomplete and there is need of some extra variables to take this into
account.  {}From the violation of Bell's inequality (see \cite{bell},
\cite{aspect}) it followed that their assumption was wrong: reality
{\it is} non local and therefore {}from their reasoning we cannot
conclude anything concerning the existence of hidden variables.

We should emphasize that the reason for introducing the configuration
of the particles as an extra variable in \QM{} has nothing to do with
nonlocality.  This has created and indeed still creates a lot of
confusion in understanding which are the consequences of the violation
of Bell's inequality---that reality is nonlocal and that any
completion of \QM{} with {\it local} hidden variable is impossible.
This is not the case of \BM{}, in which nonlocality follows directly
{}from the fact that the \wf{} is a function in configuration space,
not in ordinary space.  This means that the velocity of each particle
of a system composed by $N$ particles, independently on how far are
they.  The degree of {\it action at distance} depends on the degree of
entanglement.  It must be stressed that nonlocality is not---by any
means---a peculiarity of \BM{}: \emph{nonlocality has turned out to be
a fact of nature: nonlocality must be a feature of any physical theory
accounting for the observed violations of Bell's inequality}
\cite{bell}.

The so called ``no-go'' theorems for hidden variables (von Neumann
\cite{vNe55}, Gleason \cite{gleason}, Kochen-Speker
\cite{kochenspecher}) show that there is no ``good'' map from
operators to random variables (on the space of ``hidden variables''),
where by ``good'' we mean in the sense that the joint distributions of
the random variables are consistent with the corresponding quantum
mechanical distributions whenever the latter are defined.  As commonly
understood, these theorems involve a certain irony: They conclude with
the impossibility of a deterministic description, or more generally of
any sort of realist description, only by in effect themselves assuming
a ``realism'' of a most implausible variety, namely, naive realism
about operators \cite{daumer0}.  There is in fact no reason to expect
there to be such a map: the fact that the same operator plays a role
in different experiments does not imply that these experiments have
much else in common, and certainly not that they involve measurements
of the same thing.  It is thus with detailed experiments, and not with
the associated operators, that random variables might reasonably be
expected to be associated \cite{bell}, \cite{daumer0}, \cite{cushing}.
\medskip

Finally, it is interesting to note, as a side remark, that the true
``hidden'' variable is actually the \wf{}.  In fact, it is not
stressed sufficiently that it is indeed the \wf{} that cannot be
measured!  If the \wf{} were measurable, it would exist an
experimental device revealing the the actual wave function $\psi_{0}$
of the system prior to the measurement and the statistics of the
pointer measuring the \wf{} would be formally given by $ \d
(\psi-\psi_0)$, which, however, is not a quadratic form of $\psi_0$, so
that (\ref{meas}) is violated and and thus the \wf{} is not
measurable!

\section{What about Relativity?}\label{sec:war}

\BM, the theory defined by eqs.  (\ref{eq:bohmeq}) and
(\ref{eq:scevol}) (together with (\ref{eq:hamiltonian}) and
(\ref{eq:velofield})), is not Lorentz invariant, since
(\ref{eq:scevol}) is a nonrelativistic equation, and, more
importantly, since the right hand side of (\ref{eq:velofield})
involves the positions of the particles at a common (absolute) time.
It is also frequently asserted that \BM\ cannot be made Lorentz
invariant, by which it is presumably meant that no Bohmian theory---no
theory that could be regarded somehow as a natural extension of
\BM---can be found that is Lorentz invariant.  In this regard, we wish
to make some remarks.
\begin{enumerate}
\item The main reason for the belief that \BM\ cannot be made Lorentz
is the manifest nonlocality of \BM{}, but nonlocality, as we have
stressed in the previous section, is a fact of nature.

\item Concerning the other (somehow related) widespread belief, that
standard quantum theories have no problems incorporating relativity
while \BM{} does, we completely agree with the assessment of Jean
Bricmont: ``Indeed, whatever the Copenhagen interpretation really
means, it must somewhere introduce a collapse or reduction of the
state vector or an intervention of the observer or some---usually
verbal---substitute for that.  And, to my knowledge the Lorentz
invariance of that part of the ``theory'' is never discussed.  And, if
it was, problems caused by EPR-type experiments, that are the source
of the difficulty in obtaining a Lorentz invariant Bohmian theory,
would soon become evident.  \cite{bricmont}''

\item Indeed, the Bohmian state description $(\Psi,Q)$ has been extended
to (bosonic) field theories with $Q$ representing the instantaneous
configuration of the field (see, e.g., \cite{bohm52},
\cite{bohmhiley}, \cite{holland}).  Though this might be the
appropriate ontology for relativistic physics, it should be stressed
that Bell (\cite{bell}, 173--180) has proposed a Bohmian model for a
quantum field theory involving both bosonic and fermionic quantum
fields in which the primitive ontology is associated \emph{only} with
fermions.

\item While the above extensions agree with the predictions of quantum field
theory---and thus they are relativistically invariant on the
phenomenological level---they seem to lack ``\emph{serious}'' Lorentz
invariance (as Bell has put it \cite{bell}) on the level of the
basic dynamical laws (for a discussion of this point see
\cite{berndl1}).

\item Finally, we'd like to stress
that it is indeed possible to construct ``\emph{serious}'' Lorentz invariant
Bohmian models, i.e., models for which Lorentz invariance holds, non
only on the phenomenological level, but also on the microscopic level
of the basic dynamical laws~\cite{berndl1},~\cite{berndl2},~\cite{rod}.
\end{enumerate}

\section{Acknowledgments}

This work was financially supported in part by the INFN. Part of the
work has grown and has been developed in the IHES, the Mathematisches
Institut der Universit\"at M\"unchen and the Department of Mathematics
of the University of Rutgers.  The hospitality of these institutions
is gratefully acknowledged.  Finally, we thank Detlef D\"urr and
Shelly Goldstein for their teaching, support and friendship.

\end{document}